\documentstyle [aps,psfig]{revtex}

\begin{document}
\draft
\title{The Singularity in Generic Gravitational Collapse
Is Spacelike, Local, and Oscillatory \thanks{This essay received 
an ``honorable mention'' from the Gravity Research Foundation, 1998 -  Ed. }}
\author{B. K. Berger,$^1$ D. Garfinkle,$^1$ J. Isenberg,$^2$ V.
Moncrief,$^3$ M. Weaver$^4$}
\address{$^1$Department of Physics, Oakland University, Rochester, MI 48309 USA}
\address{$^2$Department of Mathematics, University of Oregon, Eugene, OR 97403}
\address{$^3$Departments of Physics and Mathematics, Yale University, New Haven,
CT 06520 USA}
\address{$^4$Department of Physics, University of Oregon, Eugene, OR  97403}

\maketitle
\bigskip
\begin{abstract}
A longstanding conjecture by Belinskii, Khalatnikov, and Lifshitz that the
singularity in generic gravitational collapse is spacelike, local, and
oscillatory is explored analytically and numerically in spatially inhomogeneous
cosmological spacetimes. With a convenient choice of variables, it can be seen
analytically how nonlinear terms in Einstein's equations control the approach to
the singularity and cause oscillatory behavior. The analytic picture requires
the drastic assumption that each spatial point evolves toward the singularity
as an independent spatially homogeneous universe. In every case,
detailed numerical simulations of the full Einstein evolution equations support
this assumption.
\end{abstract}

{\it Introduction.} 
Even though much of the theoretical effort in gravitational physics is
focused on either quantum gravity or astrophysical sources of gravitational
waves, the most fundamental properties of general solutions to Einstein's
equations remain unknown. Foremost among these is the nature of the singularity
that forms in generic gravitational collapse. Powerful theorems due to Penrose
and Hawking and others \cite{hawking70,wald84} predict that some kind of
singular behavior will result. However, the proofs of these theorems require
very little of the details of the field equations and are thus unable to
provide details of the nature of the singularity. Many different types of
singular behavior arise in a variety of special cases, but which, if any, of
these characterize generic collapse is not known. In fact, it is hard to see
that one could hope to answer the question. Einstein's equations are second
order nonlinear partial differential equations (PDE's) in $3 + 1$ spacetime
dimensions for the metric variables restricted by four constraints and with the
freedom to impose four coordinate conditions. It is by no means clear that
numerical relativity can help to answer this question either since a code will
``blow up'' if any variables become singular.

The surprising message of this essay then is that progress in understanding
singularities in generic gravitational collapse can indeed be made. It requires
a judicious choice of variables and a synergy between numerical simulations
which suggest the patterns and analytic predictions which guide the
simulations. Our conclusions are based on the properties of solutions with one
or two directions of spatial symmetry. While these are not the most general
solutions of Einstein's equations, we believe they are sufficiently generic and
sufficiently consistent with each other to indicate how completely general
solutions are likely to behave. 

Approximately thirty years ago, Belinskii, Khalatnikov, and Lifshitz (BKL)
\cite{belinskii71b} proposed, based on a perturbative analysis, that the
singularity in generic collapse was spacelike, local, and oscillatory. In their
picture, each spatial point evolves toward the singularity as a distinct
Mixmaster universe. Their claim has been criticized but never unambiguously
proven or refuted. In addition substantial effort is required to follow their
arguments. Our recent work, however, substantially supports their predictions
and suggests a simple scenario:  It appears that, despite their complexity,
Einstein's equations in the approach to the singularity have an attractor which
is dominated by nonlinear terms acting as effective potentials for the
dynamics. These potential terms act primarily on one of the gravitational field
degrees of freedom and lead to a generic asymptotic solution that supports the
BKL description of the generic singularity as spacelike, local, and
oscillatory. Other information, such as spatial dependence and the behavior of
the other degrees of freedom, is carried along by the dominant behavior. 

Our method of studying a given family of solutions as they approach the
singualrity is essentially the following \cite{grubisic93}:  We first determine
the velocity-term dominated (VTD) solutions corresponding to the family. These
are solutions of the ODE's which result from the truncation of Einstein's
equations to eliminate all terms containing spatial derivatives (plus certain
others; see
\cite{isenberg90}). Note that the VTD solutions are parametrized by
spatially-dependent constants of integration (the ``VTD parameters''). We then
substitute these solutions into the Einstein equations or into the corresponding
Hamiltonian. If, for generic VTD parameters, the terms in the equations or the
Hamiltonian containing spatial derivatives become exponentially small (hence,
dynamically irrelevant) as $\tau \to \infty$ ($\tau = \infty$ is the time of
the spacelike singularity), then we predict that the family is asymptotically
velocity-term dominated (AVTD) in the sense that a generic solution of the
Einstein equations asymptotically approaches a VTD solution. If not, we monitor
which terms might become important (as a result of exponential growth), and we
try to predict the subsequent behavior.

We have applied this ``Method of Consistent Potentials'' (MCP) to various
families of cosmological solutions such as the Gowdy spacetimes
\cite{gowdy71,moncrief81}, the Gowdy spacetimes with magnetic fields
\cite{weaver98}, and
$U(1)$ symmetric spacetimes \cite{moncrief86,berger97e}. In all cases, using a
choice of time based on geometric invariants connected with the symmetry, we
get a clear prediction:  either that the spacetime solutions in the family are
AVTD, or that they have oscillatory behavior characterized by a sequence of
AVTD-like epochs punctuated by ``bounces'' which involve the short-term
awakening of certain of the ``potential'' terms. Moreover, numerical
simulations of solutions of the full Einstein equations (for each family) have
in all cases agreed (at generic spatial points) with the MCP predictions. While
these numerical simulations cannot be followed all the way to the singularity,
they can be taken sufficiently far, with sufficient accuracy, to verify the
simple BKL scenario.

We shall now discuss some of the details of these studies, starting with a look
at spatially homogeneous cosmologies.

{\it Homogeneous cosmologies.} For a spatially homogeneous cosmology with flat
spacelike hypersurfaces (Bianchi I), the Hamiltonian for Einstein's theory
takes the form
\begin{equation}
\label{kasnerh}
H=-p_\Omega ^2+p_+^2+p_-^2 = 0,
\end{equation}
where $p_\Omega$, $p_\pm$ are respectively canonically conjugate to $\Omega$,
the logarithm of the isotropic scale factor, and
$\beta_\pm$, the (logarithmic) anisotropic shears, and we have the Kasner
solutions \cite{Kasner25}. These are parametrized by the constants $v_\pm =
p_\pm / | p_\Omega|$. Asymptotically ($\Omega \to -\infty$), the shear satisfies
\begin{equation}
\label{kasnersoln}
\beta_\pm = - v_\pm \Omega
\end{equation}
with $v_\pm $ satisfying as a consequence of the
constraint (\ref{kasnerh})
\begin{equation}
\label{kasnerrestrict}
v_+^2 \, + \, v_-^2 = 1.
\end{equation}

For more general homogeneous cosmologies, the Hamiltonian
$H$ adds a term $V(\Omega,\beta_\pm)$ which in certain cases \cite{ryan75} takes
the form
\begin{equation}
\label{mixmasterv}
V(\Omega,\beta_\pm) = c_1 e^{a_1 (\Omega - 2 \beta_+)} + c_2 e^{a_2 (\Omega +
\beta_+ + \sqrt{3} \beta_-)} + c_3 e^{a_3 (\Omega +
\beta_+ - \sqrt{3} \beta_-)} + \ \dots \ \ ,
\end{equation}
where $c_i$ and $a_i$ are positive constants. The solutions corresponding to
$H$ with this form of the potential are the well-studied Mixmaster solutions
\cite{belinskii71b,misner69,hobill94}. To see the oscillatory behavior of
Mixmaster solutions from the MCP point of view, we note that the Bianchi I
Hamiltonian (\ref{kasnerh}) is exactly the VTD version of the Mixmaster
Hamiltonian, and the VTD solutions are the Kasner spacetimes. If we now
substitute these Kasner solutions into (\ref{mixmasterv}), as $\Omega \to
-\infty$ we have
\begin{equation}
\label{mixmastervlimit}
V\to c_1e^{a_1|\Omega |(-1-2v_+)}+c_2e^{a_2|\Omega |(-1+v_++\sqrt
3v_-)}+c_3e^{a_3|\Omega |(-1+v_+-\sqrt 3v_-)} \equiv V_1 + V_2 + V_3.
\end{equation}
A few moments of reflection will show that the restriction
(\ref{kasnerrestrict}) implies that any generic values of $v_\pm$ will cause
one of the $V_j$ to grow and the others to decay. The growing term
provides an exponential potential for the dynamics. A bounce off the potential
changes the sign of the appropriate combination of momenta. What happens then
is that a pattern of growing and decaying terms continues indefinitely. It has
long been known that this picture is borne out by the full solutions to
Einstein's equations \cite{belinskii71b,berger96c}. An example showing the
evolution of the $V$'s and $\beta_+$ vs $|\Omega|$ is given in Fig.~1.

{\it Inhomogeneous cosmologies.} We first consider the Gowdy family of
spacetimes on $T^3 \times R$. These allow spatial dependence in one direction
with periodic boundary conditions and are described in terms of
\cite{gowdy71,moncrief81,weaver98} 
$P(\theta,\tau)$, $Q(\theta,\tau)$, essentially the $+$ and $\times$
polarizations of gravitational waves, and a background
$\lambda(\theta,\tau)$. This model has the nice feature
that the dynamical equations for the wave amplitudes $P$, $Q$ decouple from the
constraints which yield $\lambda$ as a function of the wave amplitudes.

Einstein's evolution equations are obtained from the Hamiltonian density
\cite{moncrief81,weaver98}
\begin{equation}
\label{gowdyh}
{\cal H}={1 \over {4\pi _\lambda }}\left( {\pi _P^2+e^{-2P}\pi _Q^2+e^{-2\tau
}P'^2+e^{2(P-\tau )}Q'^2} \right)
\end{equation}
where the $\pi$'s are the conjugate momenta and $' = \partial/\partial
\theta$.\footnote{In the usual treatment of this model
\cite{berger93,berger97b}, the non-dynamical
$\pi_\lambda$ is set equal to ${\textstyle{{1} \over {2}}}$.} The Hamiltonian
constraint is the equation for $\dot \lambda$ (where the overdot means
$\partial / \partial \tau$) while the momentum constraint is one for
$\lambda'$. Neglecting spatial derivatives in the evolution equations yields, in
the limit
$\tau \to \infty$, the VTD solution
\begin{equation}
\label{gowdyavtd}
P = v(\theta) \tau, \quad
Q = Q_0(\theta), \quad
\lambda = - v^2(\theta) \tau,
\end{equation}
with all the momenta dependent only on $\theta$ and $v = \pi_P \,/ \, 2
\pi_\lambda$. Substitution in (\ref{gowdyh}) shows that the only terms in
${\cal H}$ which could fail to be exponentially small are
\begin{equation}
\label{gowdyv1v2}
V_1 = \left({{\pi_Q}^2 \over {4 \pi_\lambda}} \right) \,
e^{-2P} \to \left({{\pi_Q^0}^2 \over {4 \pi_\lambda^0}} \right) \, e^{-2v \tau}
,\quad V_2 = \left( {{{Q'}^2} \over {4 \pi_\lambda}} \right) \,
e^{2(P-\tau)} \to \left( {{{Q'_0}^2} \over {4 \pi_\lambda^0}} \right) \,
e^{2(v-1)\tau} .
\end{equation}
$V_1$ will grow exponentially if $v < 0$. $V_2$ will grow if $v > 1$. Bounces
off $V_1$ and $V_2$ eventually drive $v$ into the range
$0 < v < 1$, yielding a consistent AVTD solution at almost every spatial
point \cite{berger93,berger97b}. Fig.~2 shows how this process works for
$V_1$ and $V_2$ and $P$ vs $\tau$ at a typical spatial point.

The ``magnetic Gowdy'' model is obtained by a change of topology and the
addition of a magnetic field \cite{weaver98}. The VTD solution
(\ref{gowdyavtd}) is unchanged. The Hamiltonian (\ref{gowdyh}) picks up an
additional term
\begin{equation}
\label{gowdyv3}
V_3 = 4 \,\pi_\lambda\, B^2\, e^{(\lambda + \tau)/2}
\end{equation}
where the constant $B$ measures the strength of the magnetic field and
$\lambda$ is now dynamical. Substitution of the VTD solution (\ref{gowdyavtd})
into (\ref{gowdyv3}) yields as $\tau \to \infty$
\begin{equation}
\label{gowdyv3limit}
V_3 = 4 \, \pi_\lambda^0 \,B^2 \,e^{(1-v^2)\tau / 2} .
\end{equation}
The range $0 < v < 1$ which allows $V_1$ and $V_2$ to become exponentially
small causes $V_3$ to grow. Thus, in the MCP picture, no value of $v$ leads to
AVTD behavior. Further MCP qualitative study \cite{weaver98} indicates that at
each spatial point, $V_1$, $V_2$, and $V_3$ alternately awaken, causing bounces
characteristic of local Mixmaster behavior. Numerical simulations support this
prediction with typical behavior shown in Fig.~3.

It is possible to generalize the plane symmetric Gowdy model on $T^3 \times R$
by removing one of the Killing vectors. This yields vacuum $U(1)$ symmetric
models \cite{moncrief86,berger93,berger97e}. Here, the dynamical degrees of
freedom, now called
$\varphi$ and $\omega$, no longer decouple from the background described by
$\Lambda$,
$z$, and $x$. Einstein evolution equations may be obtained from the variation of
($,_a =\partial / \partial x^a$ with spatial coordinates $u$, $v$, and time
$\tau$)
\begin{eqnarray}
\label{u1h}
H &=& \int \int du \kern 1pt dv \left( {1 \over 8}p_z^2+{1 \over 2}
e^{4z}p_x^2+{1 \over 8}p_\varphi^2+{1 \over 2}e^{4\varphi }p_\omega^2-{1 \over
2}p_\Lambda ^2+2p_\Lambda  \right) \nonumber \\
&& +e^{-2\tau } \int \int du \kern 1pt 
dv \left\{  \left( {e^\Lambda e^{ab}} \right) ,_{ab}- \left( {e^\Lambda e^{ab}}
\right) ,_a\Lambda ,_b+e^\Lambda  \right. \left[  \left( {e^{-2z}}
\right) ,_u x,_v- \left( {e^{-2z}} \right) ,_v x,_u \right] \nonumber \\
&& \left. +2e^\Lambda e^{ab}\varphi ,_a\varphi ,_b+{1 \over 2}
e^\Lambda e^{-4\varphi }e^{ab}\omega ,_a\omega ,_b \right\} \nonumber \\
&=& \int \int du \kern 1pt dv \, ({\cal H}+2p_\Lambda)
\end{eqnarray}
where $p_\zeta$ is canonically conjugate to $\zeta$, $e_{ab}(z,x)$ is the
conformal metric in the $u$-$v$ plane, and
${\cal H} = 0$ is the Hamiltonian constraint. The choice $p_\Lambda > 0$ defines
the collapsing direction
\cite{berger97e}. The VTD solution as $\tau \to \infty$ is
\begin{equation}
\label{u1avtd}
z = - v_z \tau, \ \ 
x = x_0, \ \ 
\Lambda = (2 - p_\Lambda^0)\tau, \ \ 
\varphi = - v_\varphi \tau, \ \ 
\omega = \omega_0,
\end{equation}
where $v_z$, $x_0$, $p_\Lambda^0$, $v_\varphi$, $\omega_0$, and the remaining
momenta are functions of $u$ and $v$ but constants in $\tau$. Upon substitution
of the VTD solution (\ref{u1avtd}) into (\ref{u1h}), it is seen immediately
that the Gowdy-like potential terms
\begin{equation}
\label{u1v1}
V_z = {\textstyle{1 \over 2}}\, p_x^2 \,e^{4z} \to {\textstyle{1 \over
2}}\, p_x^2 \,e^{-4v_z \tau}, \quad V_1 = {\textstyle{1
\over 2}} \,p_\omega^2 \,e^{4\varphi } \to {\textstyle{1
\over 2}} \,p_\omega^2 \,e^{-4v_\varphi \tau}
\end{equation}
become negligible if $v_z > 0$, $v_\varphi > 0$, as in Gowdy. The very
complicated terms  in (\ref{u1h}) with
spatial derivatives have a simple exponential behavior. Given $v_z >
0$, the inverse conformal metric $e^{ab}$ is dominated by $e^{-2z}$.
Examination of the terms in (\ref{u1h}) shows that all but one of them
have the exponential factor
\begin{equation}
\label{u1vlam}
e^{(-2 \tau + \Lambda - 2z)} \approx e^{(-p_\Lambda^0 +2 v_z) \tau}.
\end{equation}
The remaining term is
\begin{equation}
\label{u1v2}
V_2 = {\textstyle{1 \over 2}} e^{-2 \tau + \Lambda}e^{ab}\omega,_a \omega_b
\approx F(x,\nabla \omega) e^{(-p_\Lambda^0 + 2 v_z + 4 v_\varphi) \tau}.
\end{equation}
Now as $\tau \to \infty$, the Hamiltonian constraint, with the VTD solution
substituted in, becomes
\begin{equation}
\label{u1havtd}
{\cal H} \approx -{\textstyle{1 \over 2}} (p_\Lambda^0)^2 +2 v_z^2 + 2
v_\varphi^2 \approx 0.
\end{equation}
This restriction means that $p_\Lambda^0 > 2 v_z$ so that $p_\Lambda^0,\ p_z
>0$ implies the exponential decay of terms with the behavior of 
(\ref{u1vlam}).\footnote{Numerical simulations show that polarized $U(1)$ models
with $V_2 \equiv 0$ have the predicted AVTD behavior \cite{berger97e}.}  On the
other hand, $V_2$ will decay exponentially only if
$(p_\Lambda^0)^2 > (2 v_z + 4 v_\varphi)^2$ which is inconsistent with
(\ref{u1havtd}). Thus $V_2$ must increase exponentially if $v_z$ and $v_\varphi$
are positive. The Hamiltonian constraint restriction means that $V_2$ will grow
approximately as $e^{4 v_\varphi \tau}$ so that $V_2$ will act as a potential
for the $\varphi$ degree of freedom. A bounce will yield $v_\varphi < 0$ which
will cause $V_1$ from (\ref{u1v1}) to grow exponentially. A bounce off $V_1$
will cause the cycle to repeat. All other variables will remain consistent with
the VTD solution. Surprisingly, this extremely simple picture describes the
results of numerical simulations of the full $U(1)$ symmetric model Einstein
equations \cite{berger98}. The behavior of $V_1$, $V_2$, and $\varphi$ vs $\tau$
at a typical spatial point is shown in Fig.~4.

{\it Conclusions.} The asymptotic approach to the singularity of a sequence of
increasingly complicated cosmological spacetimes has been examined. The
VTD solution is substituted into the relevant Hamiltonian to obtain a
prediction as to whether terms not present in the VTD solution become
exponentially small. In the homogeneous Mixmaster case, we find that no matter
what the value of the VTD constants, one of three dominant exponential
potential terms will always grow. In the vacuum Gowdy case, there are two such
potentials at every spatial point and a window, $0 < v(\theta) < 1$, consistent
with AVTD behavior. Addition of a magnetic field (with a suitable change of
topology) removes this window and causes the system to oscillate indefinitely
as the three potentials grow and decay. Detailed study shows this to be truly
local Mixmaster behavior \cite{weaver98}. On the other hand, generic $U(1)$
symmetric models have oscillations between two rather than three potentials at
every spatial point although no value of $v_\varphi(u,v)$ is consistent with
the AVTD solution.

We emphasize that, in the inhomogeneous case, this simplified picture follows
from the MCP analysis if the time dependences of a variety of spatially dependent
functions are assumed to be negligible compared to the exponentials we have
singled out. In every case, this drastic assumption holds true in numerical
simulations of the full Einstein evolution equations. This means that we could
evolve toward the singularity using ODE's at every spatial point rather than
PDE's. Thus the BKL conjecture that the generic approach to the singularity is
local is very strongly supported. However, it is not yet clear whether,
generically, the further conjecture that our observed oscillations are Mixmaster
is also correct. 

For the
inhomogeneous models, the local nature of the dynamics is plausible. It is easy
to show that the horizon size goes to zero as $\tau \to \infty$ in both the
Gowdy \cite{berger74} and $U(1)$ cases. Thus, in the chosen time coordinate,
the spatial points causally decouple. Information from other sites, carried
through the spatial derivatives, ceases to propagate.

There are several implications of these results. First, the local nature of the
approach to the singularity means that our choice of cosmological boundary
conditions is merely a convenience. The indicated behavior should arise
asymptotically in any generic collapse leading to a spacelike singularity.
Second, as bounces occur at different spatial points at different times, the
amount of spatial structure in the gravitational waves will increase
\cite{kirillov87,berger97b}. Third, these results are certainly consistent with
a generic spacelike singularity (at $\tau = \infty$) although this was not known
before these studies to be a property of $U(1)$ symmetric models.

The simplicity of our picture despite the generality of the one Killing field
spacetimes suggests that further generalization is possible. Within the models
discussed here, one may for example introduce a scalar field of the type known
to suppress Mixmaster oscillations in the homogeneous case \cite{halpern87}.
Implementation of the MCP analysis leads in all classes of inhomogeneous models
studied here to the prediction that the oscillations will be suppressed. These
predictions have not yet been checked by numerical simulation.

Finally, the generalization to the zero Killing field case may be within
reach. While precise, quantitative numerical studies in
$2 + 1$ and $3 + 1$ dimensions must still come to terms with numerical
difficulties caused by the spiky features seen in the Gowdy models
\cite{berger93,berger97b}, the local nature of the collapse means that the
qualitative features can be seen even at coarse spatial resolutions.

\section*{Acknowledgements}
This work was supported in part by NSF
Grants PHY9507313 and PHY9722039 to Oakland U., PHY9308177 to the
U.~of Oregon, and PHY9503133 to Yale U. BKB thanks the Institute for Geophysics
and Planetary Physics of Lawrence Livermore National Laboratory for hospitality.
Some numerical simulations were performed at NCSA (U.~of Illinois).


\begin{figure}[bth]
\begin{center}
\makebox[4in]{\psfig{file=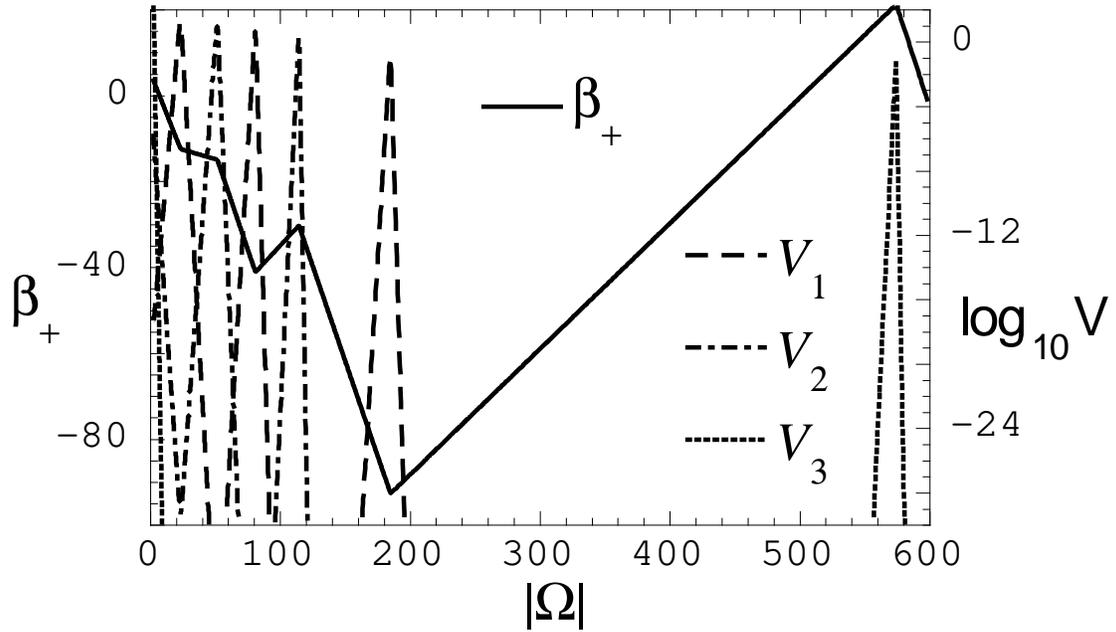,width=6in}}
\caption{A typical Mixmaster trajectory (vacuum Bianchi IX) from a numerical
simulation of Einstein's equations. The potentials $V_j$ are defined in
(\protect \ref{mixmasterv}).}
\end{center}
\end{figure}

\begin{figure}[bth]
\begin{center}
\makebox[4in]{\psfig{file=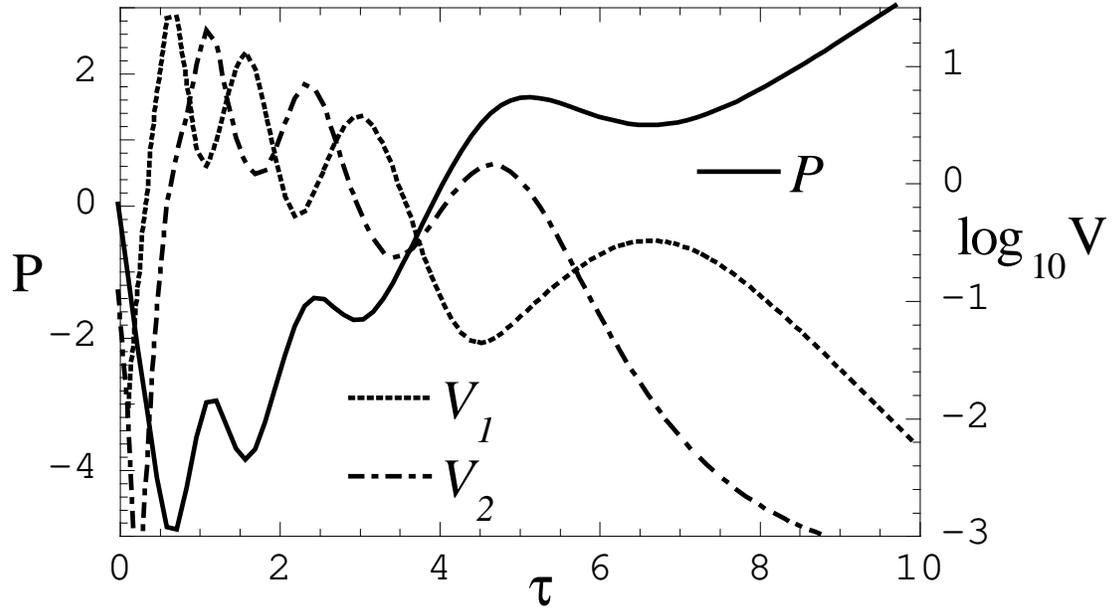,width=6in}}
\caption{Evolution at a representative spatial point in a numerical
simulation of Einstein's equations for a typical vacuum Gowdy model. The
potentials $V_1$ and
$V_2$ are defined in (\protect \ref{gowdyv1v2}).}
\end{center}
\end{figure}

\begin{figure}[bth]
\begin{center}
\makebox[4in]{\psfig{file=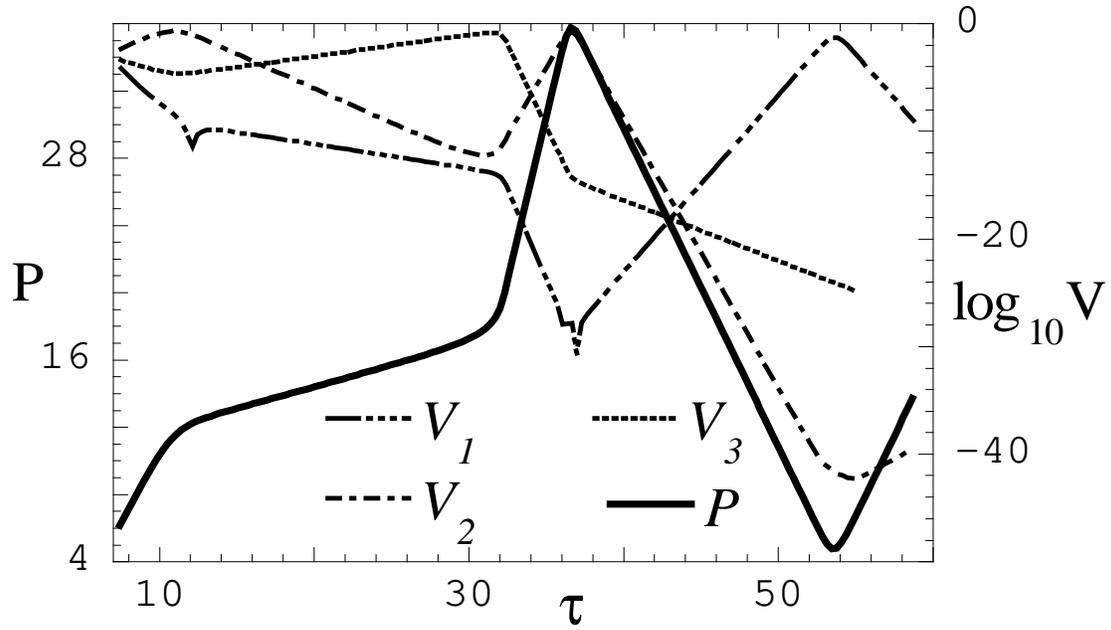,width=6in}}
\caption{Evolution at a representative spatial point in a numerical
simulation of Einstein's equations for a
typical magnetic Gowdy model. The potentials $V_1$ and $V_2$
are defined in (\protect \ref{gowdyv1v2}) while $V_3$ is defined in
(\protect \ref{gowdyv3}).}
\end{center}
\end{figure}

\begin{figure}[bth]
\begin{center}
\makebox[4in]{\psfig{file=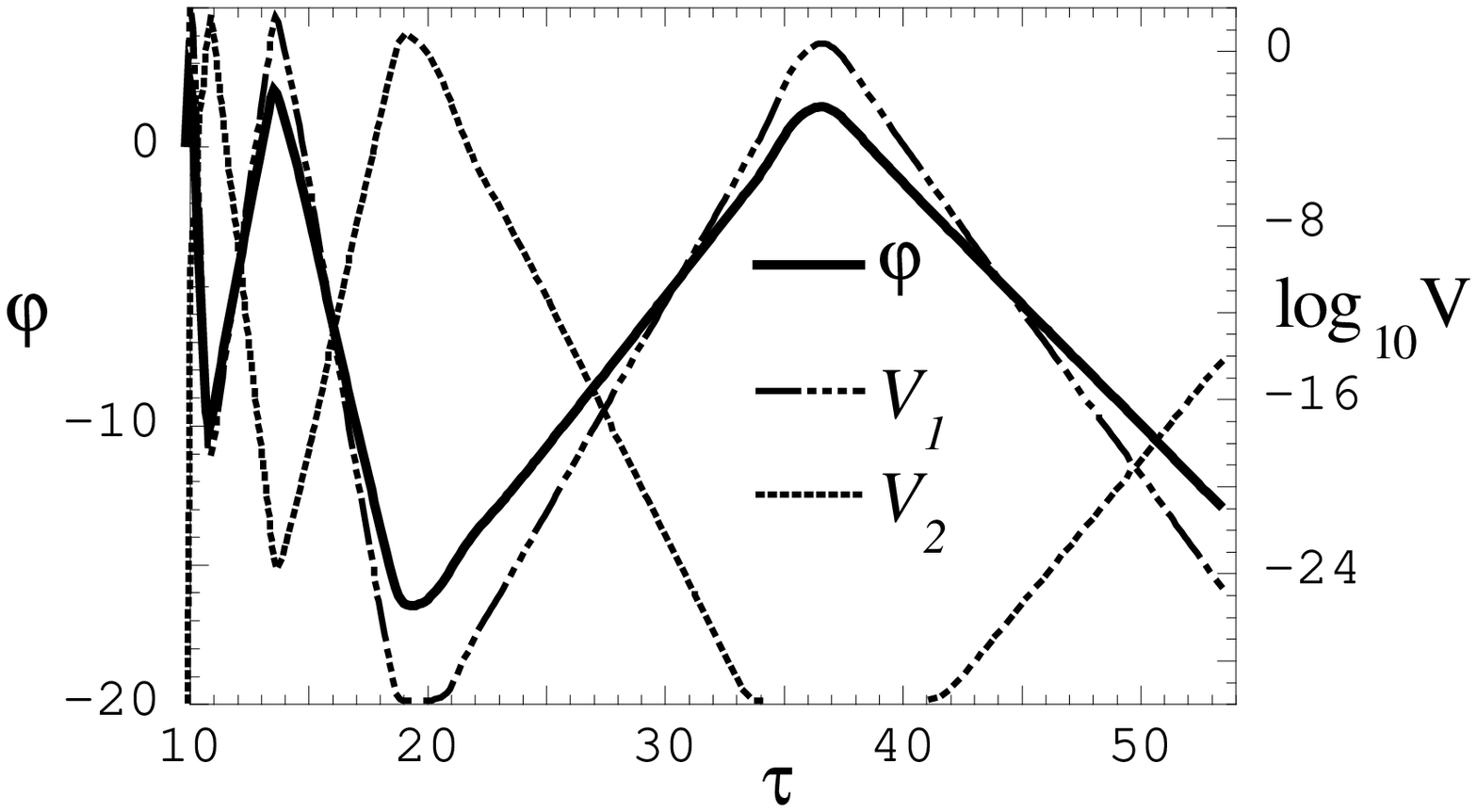,width=6in}}
\caption{Evolution at a representative spatial point in a numerical
simulation of Einstein's equations for a
typical generic $U(1)$ symmetric model. The potentials are
defined in (\protect \ref{u1v1}) and (\protect \ref{u1v2}).}
\end{center}
\end{figure}


\begin{thebibliography}{10}

\bibitem{belinskii71b}
{Belinskii, V. A.}, {Lifshitz, E. M.}, and {Khalatnikov, I.
  M.}, {\em Sov. Phys. Usp.},{\bf 13}, 745--765, (1971).

\bibitem{berger74}
Berger, B.~K., {\em Ann. Phys. (N.Y.)}, {\bf 83}, 458, (1974).


\bibitem{berger97b}
{Berger, B. K.}, and {Garfinkle, D.},
{\em Phys. Rev. D}, {\bf 57}, 4767 (1998). 

\bibitem{berger96c}
{Berger, B. K.}, {Garfinkle, D.}, and {Strasser, E.},
{\em Class. Quantum Grav.}, {\bf 14},
L29--L36, (1997). 

\bibitem{berger93}
{Berger, B. K.}, and {Moncrief, V.},
{\em Phys. Rev. D}, {\bf 48}, 4676, (1993).

\bibitem{berger97e}
{Berger, B. K.} and {Moncrief, V.},
{\em Phys. Rev. D}, {bf 57}, 15 June 1998.


\bibitem{gowdy71}
Gowdy, R.~H., {\em Phys. Rev. Lett.}, {\bf
  27}, 826, (1971). 

\bibitem{grubisic93}
{Grubi\u{s}i\'{c}, B.}, and {Moncrief, V.},
{\em Phys. Rev. D}, {\bf 47},
2371--2382, (1993). 

\bibitem{halpern87}
Halpern, P., {\em J. Gen.
  Rel. Grav.}, {\bf 19}, 73--94,
  (1987). 

\bibitem{hawking70}
{Hawking, S. W.}, and {Penrose, R.}, 
{\em Proc. Roy. Soc. Lond. A}, {\bf 314},
529--548, (1970). 

\bibitem{hobill94}
{Hobill, D.}, {Burd, A.}, and {Coley, A.}, eds.,
{\em Deterministic Chaos in General Relativity},
  (Plenum, New York, 1994).

\bibitem{isenberg90}
{Isenberg, J. A.}, and {Moncrief, V.},
{\em Ann.
  Phys. (N.Y.)}, {\bf 199}, 84,
  (1990). 


\bibitem{Kasner25}
Kasner, E., {\em
  Trans. Am. Math. Soc.}, {\bf 27}, 155--162,
  (1925). 

\bibitem{kirillov87}
{Kirillov, A. A.}, and {Kochnev, A. A.},
{\em JETP Lett.}, {\bf 46}, 435--438, (1987).



\bibitem{misner69}
Misner, C.~W., {\em Phys. Rev. Lett.}, {\bf 22},
1071--1074, (1969). 

\bibitem{moncrief81}
Moncrief, V., {\em Ann.
  Phys. (N.Y.)}, {\bf 132}, 87--107,
  (1981). 

\bibitem{moncrief86}
Moncrief, V., {\em Ann. Phys. (N.Y.)}, {\bf 167},
118, (1986). 

\bibitem{ryan75}
{Ryan, Jr., M. P.}, and {Shepley, L. C.}, {\em
  Homogeneous Relativistic Cosmologies}, (Princeton
  University, Princeton, 1975).

\bibitem{wald84}
Wald, R.~M., {\em General Relativity},
  (University of Chicago Press, Chicago,
 1984). 

\bibitem{weaver98}
{Weaver, M.}, {Isenberg, J.}, and {Berger, B. K.},
{\em Phys. Rev. Lett.}, {\bf 80}, 2984 (1998).

\bibitem{berger98}
Berger, B. K. and Moncrief, V., submitted to {\em Phys. Rev. D}, gr-qc/9804085.

\end{thebibliography}
\end{document}